\documentclass[letterpaper]{article}
\usepackage{aaai21}
\usepackage{times}
\usepackage{helvet}
\usepackage{courier}
\usepackage{xcite}
\usepackage{amsmath,amssymb,amsfonts}
\usepackage{algorithmic}
\usepackage{graphicx}
\usepackage{multirow}
\usepackage{bigstrut}
\usepackage{enumitem}
\usepackage[skip=0pt]{subcaption}
\usepackage[skip=0pt]{caption}
\usepackage{xcolor}
\usepackage{natbib}
\usepackage{bm}
\usepackage[ruled]{algorithm2e}
\usepackage{makecell}
\usepackage{tikz}

\title{Knowledge-Enhanced Top-K Recommendation in Poincaré Ball}
\author{
Chen Ma \thanks{Work done as interns at Huawei Noah’s Ark Lab in Montreal.}\textsuperscript{\rm 1}{\normalfont,}
Liheng Ma \footnotemark[1]\textsuperscript{\rm 1}{\normalfont,}
Yingxue Zhang \textsuperscript{\rm 2}{\normalfont,}
Haolun Wu \footnotemark[1]\textsuperscript{\rm 1}{\normalfont,}
Xue Liu \textsuperscript{\rm 1}
{\normalfont and} Mark Coates \textsuperscript{\rm 1} \\
}
\affiliations {
    \textsuperscript{\rm 1} McGill University,
    \textsuperscript{\rm 2} Huawei Noah's Ark Lab Montreal \\
    chen.ma2@mail.mcgill.ca, liheng.ma@mail.mcgill.ca, yingxue.zhang@huawei.com,\\ haolun.wu@mail.mcgill.ca, xueliu@cs.mcgill.ca, mark.coates@mcgill.ca
}

\begin{document}
% \title{Knowledge-Enhanced Top-K Recommendation in Poincaré Ball}
% \author{Paper ID: 5365
% }
\maketitle
\begin{abstract}
\begin{quote}
Personalized recommender systems are increasingly important as more content and services become available and users struggle to identify what might interest them. Thanks to the ability for providing rich information, knowledge graphs (KGs) are being incorporated to enhance the recommendation performance and interpretability. To effectively make use of the knowledge graph, we propose a recommendation model in the hyperbolic space, which facilitates the learning of the hierarchical structure of knowledge graphs. Furthermore, a hyperbolic attention network is employed to determine the relative importances of neighboring entities of a certain item. In addition, we propose an adaptive and fine-grained regularization mechanism to adaptively regularize items and their neighboring representations. Via a comparison using three real-world datasets with state-of-the-art methods, we show that the proposed model outperforms the best existing models by 2-16\% in terms of NDCG@K on Top-K recommendation.

\end{quote}
\end{abstract}

\section{Introduction}
With the rapid growth of Internet services and mobile devices, personalized recommender systems play an increasingly important role in modern society. They can reduce information overload and help satisfy diverse service demands. Such systems bring significant benefits to at least two parties. They can: (i) help users easily discover products from millions of candidates, and (ii) create opportunities for product providers to increase revenue. 

To provide a more accurate and interpretable recommendation service, knowledge graphs (KGs) are being incorporated into recommender systems. A KG is a heterogeneous graph, where nodes function as entities and edges represent relations between the entities. This is an effective data structure to model relational data, e.g., two movies directed by the same director. Several recent works have integrated KGs into the recommendation model, and the approaches can be divided into two branches: path-based~\cite{DBLP:conf/kdd/HuSZY18,DBLP:conf/cikm/WangZWZLXG18} and regularization-based~\cite{DBLP:conf/kdd/ZhangYLXM16,DBLP:conf/kdd/Wang00LC19}. Path-based methods extract paths from the KG that carry the high-order connectivity information and feed these paths into the predictive model. To handle the large number of paths between two nodes, researchers have either applied path selection algorithms to select prominent paths or defined meta-path patterns to constrain the paths. By contrast, regularization-based methods devise additional loss terms that capture the KG structure and use these to regularize the recommender model learning.

Although many effective models have been proposed, we argue that there are still several avenues for enhancing performance. First, previous works learn the KG representations in the Euclidean space. As has been observed in other application domains, this may not effectively capture the hierarchical structure that is known to exist within KGs~\cite{DBLP:conf/icml/SalaSGR18}. Second, methods like CKE~\cite{DBLP:conf/kdd/ZhangYLXM16}, CFKG~\cite{DBLP:journals/algorithms/AiACZ18}, and RippleNet~\cite{DBLP:conf/cikm/WangZWZLXG18} do not distinguish between neighboring entities, adjusting according to their relative importance and informativeness, when learning the representation of each entity. 
This may lead to undesirable blurring of information from relations in the KG and an incomplete understanding of an entity. Third, all the regularization-based methods adopt a fixed hyper-parameter. We argue that the regularization degree should be adaptive, taking on different values for different entities according to the relevance and value of the information from the knowledge graph. Furthermore, different training phases may need different magnitudes of regularization power values, so the hyper-parameter values should evolve during training.

To tackle the aforementioned problems, we propose a knowledge-enhanced recommendation model in the hyperbolic space, namely \textit{Hyper-Know}, to tackle the top-K recommendation task. In particular, we map the entity and relation embeddings of the KG as well as user and item embeddings to the Poincaré ball model. This allows us to capture the hierarchical structure in the KG. We incorporate an attention model in the hyperbolic space, and use the Einstein midpoint for aggregation, in order to form a representation of the neighborhood of each item in the knowledge graph. We then use a regularization term to encourage the representation of an item to remain close to the representation of its neighborhood (in the hyperbolic space). This transfers the relational and structural information from the knowledge graph to the recommendation model. 
To adaptively control the regularization effect, we model the learning of adaptive and fine-grained regularization factors as a bilevel (inner and outer) optimization problem~\cite{DBLP:journals/tec/SinhaMD18}. We build a proxy function to explicitly link the learning of the regularization related parameters with the outer objective function.
We extensively evaluate our model on three real-world datasets, comparing it with many state-of-the-art methods using a variety of performance validation metrics. The experimental results not only demonstrate the improvements of our model over other baselines but also show the effectiveness of the proposed components.

To summarize, the major contributions of this paper are:
\begin{itemize}[leftmargin=*]
\item To model the hierarchical structure of KG, we map the entity and relation embeddings of the KG into the Poincaré ball along with user and item embeddings. To the best of our knowledge, ours is the first work to consider knowledge-enhanced recommendation in the hyperbolic space.
\item To transfer the knowledge from the KG to the recommendation model, we incorporate hyperbolic attention and use the Einstein midpoint to aggregate the neighboring entities of an item to form a neighborhood representation.
\item To learn the adaptive regularization factors, we cast the learning process as a bilevel optimization problem and build a proxy function to explicitly update the regularization-related parameters.
\item Experiments on three real-world datasets show that Hyper-Know significantly outperforms the state-of-the-art methods for the top-K recommendation task.
\end{itemize}

\section{Related Work}

\subsection{General Recommendation}
Early recommendation studies largely focused on explicit feedback~\cite{DBLP:conf/www/SarwarKKR01,DBLP:conf/kdd/Koren08}. The recent research focus is shifting towards implicit data~\cite{DBLP:conf/cikm/TranLL018}. Collaborative filtering (CF) with implicit feedback is usually treated as a Top-K item recommendation task, where the goal is to recommend a list of items to users that users may be interested in. It is more practical and challenging~\cite{DBLP:conf/icdm/PanZCLLSY08}, and accords more closely with the real-world recommendation scenario. Early works mostly rely on matrix factorization techniques~\cite{DBLP:conf/icdm/HuKV08,DBLP:conf/uai/RendleFGS09} to learn latent features of users and items. Due to their ability to learn salient representations, (deep) neural network-based methods~\cite{DBLP:conf/www/HeLZNHC17,DBLP:conf/icdm/SunZMCGTH19,DBLP:conf/kdd/MaKL19} are also adopted. Autoencoder-based methods~\cite{DBLP:conf/wsdm/WuDZE16,DBLP:conf/cikm/MaZWL18,DBLP:conf/wsdm/MaKWWL19} have also been proposed for Top-K recommendation. In~\cite{DBLP:conf/kdd/LianZZCXS18,DBLP:conf/ijcai/XueDZHC17}, deep learning techniques are used to boost the traditional matrix factorization and factorization machine methods. Recently, some methods are also conducted in the hyperbolic space. HyperML~\cite{DBLP:conf/wsdm/TranT0CL20} conducts metric learning in the hyperbolic space and outperforms Euclidean counterparts. \citet{DBLP:conf/sigir/FengTCCLL20} propose to tackle the next Point-of-Interest recommendation task in the hyperbolic space.

\subsection{Knowledge Graph Enhanced Recommendation}
Knowledge graphs (KGs) are an important means to represent side information of recommender systems and have proven to be helpful to improve the recommendation performance. For example, \citet{DBLP:conf/kdd/ZhangYLXM16} propose to apply the TransR method~\cite{DBLP:conf/aaai/LinLSLZ15} to learn the KG representation as well as the item embeddings in the KG. \citet{DBLP:journals/algorithms/AiACZ18} integrate users and items with the KG and jointly learn the recommendation and KG part. \citet{DBLP:conf/www/WangZZLXG19} propose a multi-task feature learning approach for knowledge graph enhanced recommendation, where these two parts are connected with a cross-and-compress unit to transfer knowledge and share regularization of items. Another track of research tries to perform propagation over the KG to assist in recommendation. Specifically, RippleNet~\cite{DBLP:conf/cikm/WangZWZLXG18} extends the user’s interests along KG links to discover her potential interests by introducing preference propagation, which automatically propagates users’ potential preferences and explores their hierarchical interests in the KG.
KPRN~\cite{DBLP:conf/aaai/WangWX00C19} constructs the extracted path sequence with both the entity embedding and the relation embedding. These paths are encoded with an LSTM layer and the preferences for items in each path are predicted through fully-connected layers. KGCN~\cite{DBLP:conf/www/WangZXLG19} studies the utilization of Graph Convolutional Networks (GCNs) for computing embeddings of items via propagation among their neighbors in the KG. Recently, KGAT~\cite{DBLP:conf/kdd/Wang00LC19} recursively performs propagation over the KG via a graph attention mechanism that refines entity embeddings. Several subsequent works~\cite{DBLP:conf/sigir/ChenZMLM20,DBLP:conf/www/WangX000C20} focus on optimizing the negative sampling procedure in knowledge-enhanced recommendation. In this paper, we report results for our proposed method using a vanilla negative sampling strategy, so that we can focus on the performance impact of the novel aspects: learning in the hyperbolic space, using hyperbolic attention with Einstein midpoint aggregation, and introducing adaptive regularization. But the advanced negative sampling strategies can also be incorporated into our proposed method to provide a further performance improvement.

% However, our proposed model is different from previous models. We learn the knowledge-enhanced recommendation in the Poincaré Ball model. 
Our proposed model distinguishes itself from previous models by learning knowledge-enhanced recommendation in the Poincaré ball model.
In addition, we employ a hyperbolic attention model in the hyperbolic space to assign different degrees of importance to the neighboring entities of a certain item. We introduce a bilevel optimization formulation of the learning task to achieve an adaptive regularization mechanism that controls the regularization effect.

\section{Preliminaries}
\subsection{Problem Formulation}
The knowledge-based recommendation considered in this paper takes as inputs the user implicit feedback and the item knowledge graph. The implicit feedback is represented by a number of user-item pairs $ \mathcal{D} = {(u, v)} \subseteq \mathcal{U} \times \mathcal{I} $, where $ \mathcal{U} $ is the user set and $ \mathcal{I} $ is the item set. 
% For simplicity, we use $ \mathbf{u} \in \mathbb{R}^{d} $ and $ \mathbf{v} \in \mathbb{R}^{d} $ to denote the hidden representation of users and items, respectively. 
The item knowledge graph $ \mathcal{G} = {(h, r, t)} \subseteq \mathcal{E} \times \mathcal{R} \times \mathcal{E}$ can be formulated as a set of triples, each consisting of a relation $ r \in \mathcal{R} $ and two entities $ h,t \in \mathcal{E} $, referred to as the \textit{head} and \textit{tail} of the triple. 
% For simplicity, we use $ \mathbf{e} \in \mathbb{R}^{d} $ and $ \mathbf{r} \in \mathbb{R}^{d} $ to denote the hidden representation of entities and relations, respectively. 

Then the top-$ K $ recommendation task in this paper is formulated as: given the training item set $ \mathcal{S}_{u} $ of user $ u $, and the non-empty test item set $ \mathcal{T}_{u} $ (requiring that $ \mathcal{S}_{u} \cup \mathcal{T}_{u} = \mathcal{D}_{u} $ and $ \mathcal{S}_{u} \cap \mathcal{T}_{u} = \emptyset $) of user $ i $, the model must recommend an ordered set of items $ \mathcal{X}_{u} $ such that $ |\mathcal{X}_{u}| \leq K $ and $ \mathcal{X}_{u} \cap \mathcal{S}_{u} = \emptyset $. Then the recommendation quality is evaluated by a matching score between $ \mathcal{T}_{u} $ and $ \mathcal{X}_{u} $, such as Recall@$K$.

\subsection{Hyperbolic Geometry of the Poincaré Ball}
The Poincaré ball model is one of five isometric models of hyperbolic geometry~\cite{cannon1997hyperbolic}, which is a non-Euclidean geometry with constant negative curvature. Formally, the Poincaré ball \(\left(\mathbb{B}_{c}^{d}, g^{\mathbb{B}}\right)\) of radius \(1 / \sqrt{c}, c>0\) is a \(d\) -dimensional manifold \(\mathbb{B}_{c}^{d}=\left\{\mathbf{x} \in \mathbb{R}^{d}: c\|\mathbf{x}\|^{2}<1\right\}\) equipped with the Riemannian metric \(g^{\mathbb{B}}\) which is conformal to the Euclidean metric \(g^{\mathbb{E}}=\mathbf{I}_{d}\) with the conformal factor \(\lambda_{\mathbf{x}}^{c}=2 /\left(1-c\|\mathbf{x}\|^{2}\right)\), i.e., \(g^{\mathbb{B}}=\left(\lambda_{\mathbf{x}}^{c}\right)^{2} g^{\mathbb{E}} .\) The distance between two points \(\mathbf{x}, \mathbf{y} \in \mathbb{B}_{c}^{d}\) is measured along a geodesic (i.e. a shortest path between the points) and is given by:
\begin{equation}
    d_{\mathbb{B}}(\mathbf{x}, \mathbf{y})=\frac{2}{\sqrt{c}} \tanh ^{-1}\left(\sqrt{c}\left\|-\mathbf{x} \oplus_{c} \mathbf{y}\right\|\right) \,,
\label{eq:hyper_dist}
\end{equation}
where $ \|\cdot\| $ denotes the Euclidean norm and $ \oplus_{c} $ represents Möbius addition~\cite{DBLP:conf/nips/GaneaBH18}:
\begin{equation}
    \mathbf{x} \oplus_{c} \mathbf{y}=\frac{\left(1+2 c\langle\mathbf{x}, \mathbf{y}\rangle+c\|\mathbf{y}\|^{2}\right) \mathbf{x}+\left(1-c\|\mathbf{x}\|^{2}\right) \mathbf{y}}{1+2 c\langle\mathbf{x}, \mathbf{y}\rangle+c^{2}\|\mathbf{x}\|^{2}\|\mathbf{y}\|^{2}} \,.
\label{eq:mobius_add}
\end{equation}

\begin{figure*}[ht!]
    \centering
    % \begin{minipage}{.3\textwidth}
    \begin{tikzpicture}
    % nodes:
    % node styles
   \tikzstyle{rec} = [rectangle, draw=black!100, fill=white!20, align=left];
    \tikzstyle{rrec} = [rectangle, draw=black!100, fill=white!20, align=center, rounded corners=2.5];
    \tikzstyle{grrec} = [rectangle, draw=green!50!black, fill=white!20, align=center, rounded corners=2.5];
    \tikzstyle{rrrec} = [rectangle, draw=red!50!black, fill=white!20, align=center, rounded corners=2.5];
% edge styles
    \tikzstyle{to} = [-latex, draw, very thick];
    \tikzstyle{gto} = [-latex, draw=green!50!black, very thick];
     \tikzstyle{rto} = [-latex, draw=red!50!black, very thick];

% Main architecture
%% nodes

\node (emb) at (0, 2) [grrec] {User \& Item \\ Embeddings \\ 
$\bm{\Theta}$};
\node (beta) at (0, 0) [rrrec] {KG Control \\ Parameters \\ $\bm{\beta}$};
% \node (lr) at (2, 2.5) [rrec] {$\mathcal{L}_R$}; 
% \node (lk) at (2, 1.5) [rrec] {$\mathcal{L}_K$};
\node (inner) at (6, 2) [rrec] {$\mathcal{J}_{inner}(\bm{\Theta}, \bm{\beta})$};
\node (proxy) at (6, 0.3) [rrec] {Proxy Function \\ \scalebox{0.8}{$\tilde\Theta(\bm{\beta}) := \Theta - \alpha \nabla_{\Theta} \mathcal{J}_{inner}(\Theta, \bm{\beta})$}};
\node (outer) at (10, 0.3) [rrec] {$\mathcal{J}_{outer}\big(\tilde\Theta(\bm{\beta})\big)$};
\node (legend) at (10,2) [rec] {\textcolor{black}{$\bm\to$} \ forward pass \\
\textcolor{green!50!black}{$\bm\to$} \ gradient backward (inner loss) \\
\textcolor{red!50!black}{$\bm\to$} \ gradient backward (outer loss)};

%% edges
\path [->, draw, thick] (emb) edge (inner);
\path [-, draw, thick] (beta) -| (2.5, 0) -- (2.5, 2);
\path [<-, draw=green!50!black, thick, transform canvas={yshift=0.15cm}] (emb) edge node[above, black!50!green, transform canvas={yshift=0.1cm, xshift=0.5cm}]{\scalebox{0.8}{$\Theta \leftarrow \operatorname{OPT}_{\Theta}\left(\Theta, \nabla_{\Theta} \mathcal{J}_{inner}(\Theta, \bm{\beta})\right)$}} (inner);

\path [->, draw, thick, transform canvas={xshift=-0.075cm}] (inner) -- (proxy);
\path [->, draw, thick, transform canvas={yshift=0.075cm}] (proxy) -- (outer);
\path [<-, draw=red!50!black, thick, transform canvas={yshift=-0.075cm}] (proxy) -- (outer);
\path [<-, draw=red!50!black, thick, transform canvas={xshift=0.075cm}] (inner) -- (proxy);
\path [-, draw=red!50!black, thick, transform canvas={yshift=-0.15cm}] (inner) -- (2.65, 2);
\path [<-, draw=red!50!black, thick, transform canvas={yshift=-0.15cm}] (beta) -| node[below, red!50!black, transform canvas={xshift=0.5cm, yshift=0.1cm}]{\scalebox{0.8}{$\bm{\beta} \longleftarrow \operatorname{OPT}_{\bm{\beta}} \Big( \bm{\beta}, \nabla_{\bm{\beta}}\mathcal{J}_{outer}\big(\tilde\Theta(\bm{\beta})\big)\Big)$}}  (2.65, 0) -- (2.65, 2);
    \end{tikzpicture}
    \vspace{0.4cm}
    % \end{minipage}
    \caption{The procedure of the bilevel optimization to control the learning of fine-grained parameters $\bm\beta$.}
    \label{fig:my_label}
\vspace{-0.3cm}
\end{figure*}

\section{Methodology}
In this section, we introduce the proposed model, \textit{Hyper-Know}, which integrates the knowledge graph with the recommendation task in the hyperbolic space. We first introduce the user preference learning in the hyperbolic space. Then we illustrate the hyperbolic attention mechanism that is used to distinguish items' neighboring entities in the knowledge graph. We next explain how to adaptively learn the recommendation objective and knowledge graph by a bilevel optimization formulation. 
Lastly we introduce the training and prediction procedure of the proposed model.

\subsection{Learning User Preference}
User preference modeling lies at the core of recommender systems. Recently, distance metric learning has been widely applied to measure the user preference on items, yielding substantial performance gains~\cite{DBLP:conf/www/HsiehYCLBE17}. In this approach, the distance $ d_{\mathbb{B}}(\mathbf{u}, \mathbf{v}) $ between user $ u $ and item $ v $ is used to measure the user preference on a certain item. To learn the user preference, we apply the Bayesian Personalized Ranking loss~\cite{DBLP:conf/uai/RendleFGS09} to capture the pairwise preference of a user $u$ for an item $ v $ that the user has accessed compared to a randomly sampled item $ {v}' $:
% \begin{equation}
%     \mathcal{L}_{R} = \sum_{(u, v) \in \mathcal{D} \wedge (u, {v}') \not\in \mathcal{D}} - \mathrm{ln} \, \sigma \left(d_{\mathbb{B}}(\mathbf{u}, {\mathbf{v}}') - d_{\mathbb{B}}(\mathbf{u}, \mathbf{v}) \right) \,,
% \label{eq:bpr_dist}
% \end{equation}
\begin{equation}
    \mathcal{L}_{R}(u, v, {v}'; \Theta) = - \mathrm{ln} \, \sigma \left(d_{\mathbb{B}}(\mathbf{u}, {\mathbf{v}}') - d_{\mathbb{B}}(\mathbf{u}, \mathbf{v}) \right) \,,
\label{eq:bpr_dist}
\end{equation}
where $ \mathbf{u} $, $ \mathbf{v} $, and $  {\mathbf{v}}' \in \mathbb{B}_{c}^{d} $, $ \sigma $ is the sigmoid function, and $ d $ is the dimension of the manifold. $\Theta$ represents the parameters of the recommender model.

\subsection{Regularizing Neighboring Entities}
Knowledge graphs (KGs), consisting of (head entity, relationship, tail entity) triples, are efficient data structures for representing factual knowledge and are widely used in applications such as question answering~\cite{DBLP:conf/aaai/ZhangDKSS18}. Recently, KGs have been applied in recommender systems to not only enhance the recommendation performance but also provide interpretable recommendation results.

To effectively exploit KGs in recommender systems, we treat them as relational inductive biases~\cite{DBLP:journals/corr/abs-1806-01261} between items. During the learning process, the relations in the KG can be used as regularizers; if two items link to a common entity or multiple common entities in the KG which suggests that a user might have similar preferences for the items. However, an item can link to multiple entities in the KG and the relative importance of different entities can differ greatly. Moreover, the entities can contribute in different ways to the description of the item. This motivates us to propose an attention mechanism in the Poincaré ball model.

Considering an item $ v $, we use $ \mathcal{N}_v = \{ (v, r, t)| (v, r, t) \in \mathcal{G} \} $ to denote the set of neighboring triples for which $ v $ is the head entity. Then we apply a TransE-style~\cite{DBLP:conf/nips/BordesUGWY13} scoring function ($ ||\mathbf{h} + \mathbf{r} - \mathbf{t}|| $) to calculate the matching score between an item and its neighboring entity in $ \mathcal{N}_v $:
\begin{equation}
    \alpha(v, t) = \mathrm{exp}\big(-d_{\mathbb{B}}(\mathbf{v}  \oplus_{c} \mathbf{r}, \mathbf{t}) \big) \,.
\end{equation}

The usual way to aggregate multiple attentions in the Euclidean space is weighted midpoint aggregation. The corresponding operation in the hyperbolic space is not immediately obvious, but fortunately, the extension to hyperbolic space does exist in the form of the \textit{Einstein midpoint}. It has a simple form in the Klein disk model~\cite{cannon1997hyperbolic}:
\begin{equation}
\label{eq:neighbourhood}
    \mathbf{n}_v = f_{\mathbb{K} \rightarrow \mathbb{B}} \left( \sum_{t\in\mathcal{N}_v} \frac{\alpha(v,t) \gamma(\mathbf{t})}{\sum_{{t}'} \alpha(v, {t}') \gamma({\mathbf{t}}')} f_{\mathbb{B} \rightarrow \mathbb{K}} (\mathbf{t}) \right) \,,
\end{equation}
where the elements of $ \gamma(\mathbf{t}) = \frac{1}{\sqrt{1 - c||\mathbf{t}||^{2}}}$ are the Lorentz factors and $ f_{\mathbb{B} \rightarrow \mathbb{K}} (\cdot)$ is the function to transform the coordinates from the Poincaré ball model to the Klein disk model. The Klein model is supported on the same space as the Poincaré ball, but the same point has different coordinates in each model. Let $ \mathbf{x}_{\mathbb{B}} $ and $ \mathbf{x}_{\mathbb{K}} $  denote the coordinates of the same point $\mathbf{x}$ in the Poincare and Klein models correspondingly. Then the following transition formulas hold:
\begin{equation}
\begin{aligned}
    \mathbf{x}_{\mathbb{K}} &= f_{\mathbb{B} \rightarrow \mathbb{K}} (\mathbf{x}_{\mathbb{B}}) = \frac{2 \mathbf{x}_{\mathbb{B}}}{1 + c||\mathbf{x}_{\mathbb{B}}||^2} \,, \\
    \mathbf{x}_{\mathbb{B}} &= f_{\mathbb{K} \rightarrow \mathbb{B}} (\mathbf{x}_{\mathbb{K}}) = \frac{\mathbf{x}_{\mathbb{K}}}{1 + \sqrt{1 - c||\mathbf{x}_{\mathbb{K}}||^2}} \,.
\end{aligned}
\end{equation}

We call $\mathbf{n}_v$ in~\eqref{eq:neighbourhood} the neighborhood representation of item $ v $. During the training process we add a regularizing term that encourages the neighborhood representation $\mathbf{n}_v$ to be close to the item's representation $ \mathbf{v} $.
The goal is to transfer the inductive bias in KG to the item representation:
% [I DON'T REALLY UNDERSTAND THE TERM UNDER THE SUMMATION HERE - $u$, $r$ and $t$ don't appear in the expression are we summing over all $v$ such that there exist $u$, $r$, $t$ so that these expressions hold? OR DO WE IMPOSE THE SAME PENALTY MULTIPLE TIMES FOR EVERY TRIPLE CONTAINING $v$?]
% \begin{equation}
%     \mathcal{L}_{\mathcal{K}} = \sum_{(u, v) \in \mathcal{D} \wedge (v, r, t) \in \mathcal{G}} d_{\mathbb{B}}(\mathbf{v}, \mathbf{n}_v) \,.
% \end{equation}
\begin{equation}
    \mathcal{L}_{K} (v; \Theta) = d_{\mathbb{B}}(\mathbf{v}, \mathbf{n}_v) \,.
\end{equation}

Combining with the user preference learning objective $ \mathcal{L}_{R} $, the overall knowledge-enhanced objective can be:
\begin{equation}
    \mathcal{L} (u, v, {v}'; \Theta) = \mathcal{L}_{R} (u, v, {v}'; \Theta) + \beta \mathcal{L}_{K} (v; \Theta) \,,
\label{eq:overall_obj}
\end{equation}
where $ \beta $ is to balance the effect from the KG.

\subsection{Adaptive and Fine-grained Regularization}
Previous works~\cite{DBLP:conf/kdd/Wang00LC19,DBLP:conf/cikm/WangZWZLXG18,DBLP:conf/kdd/ZhangYLXM16} that derive information from a KG in the recommender setting use a single and fixed number for $ \beta $ in Eq.~\ref{eq:overall_obj} to train the overall objective. However, employing a single fixed value for $ \beta $ can have several drawbacks. First, different datasets may require different impact levels of regularization from KGs. Treating $\beta$ as a fixed value requires extra hyper-parameter searching procedure for each dataset to better realize the power of KGs.
Second, different items may need different degrees of regularization. Using the same value for every item would limit the achievable performance improvement that can be derived from the KG information. Third, in different training phases, the model may need different magnitudes of regularization power. 

To address the problems outlined above, we propose an adaptive regularization scheme to apply different strengths of regularization to each item and to adjust the strength throughout training. We formulate Eq.~\ref{eq:overall_obj} as:
\begin{equation}
\begin{aligned}
    \mathcal{L} (u, v, {v}'; \Theta, \bm{\beta}) = \mathcal{L}_{R} (u, v, {v}'; \Theta) + \sigma(\beta_v) \mathcal{L}_{K} (v; \Theta) \,,
 \end{aligned}
\end{equation}
where $ \beta_v $ is $v$-th value of $ \bm{\beta} \in \mathbb{R}^{|\mathcal{I}|} $ and $ \sigma(\beta_v) \in (0, 1) $ where $ \sigma(\cdot) $ is the sigmoid function. Unfortunately, directly minimizing this objective function is not able to achieve the desired purpose of adaptively controlling the regularization. The reason is that, considering $ \beta_v $ explicitly appears in the loss function, constantly decreasing the value of $ \beta_v $ is the straightforward way to minimize the loss. As a consequence, instead of reaching optimal values for the model, all $ \beta_v $ will end up with very small values close to zero, leading to unsatisfactory results.

To tackle the above problem, we model the learning of recommendation models and the adaptive regularization of KG as a bilevel optimization problem~\cite{DBLP:journals/anor/ColsonMS07}:
\begin{equation}
\begin{aligned}
    \underset{\bm{\beta}}{\mathrm{min}} \: \mathcal{J}_{outer}\left(\Theta^*(\bm{\beta})\right) &:= \sum_{(u, v) \in \mathcal{D} \wedge (u, {v}') \not\in \mathcal{D}} \mathcal{L}_{R} \big(u, v, {v}'; \Theta^{*}(\bm{\beta}) \big) \\
    \mathrm{s.t.} \: \Theta^{*}(\bm{\beta}) & = \underset{\Theta}{\mathrm{argmin}}\: \mathcal{J}_{inner}(\Theta, \bm{\beta}) \\ &:= \sum_{(u, v) \in \mathcal{D} \wedge (u, {v}') \not\in \mathcal{D}} \mathcal{L} (u, v, {v}'; \Theta, \bm{\beta}) \,.
\end{aligned}
\label{eq:bilevel}
\end{equation}
Here $ \Theta $ contains the model parameters $ \mathbf{u} $, $ \mathbf{v} $, $ \mathbf{r} $ and $ \mathbf{t} $.
The objective function $\mathcal{J}_{inner}$ attempts to minimize $\mathcal{L}$ with respect to $ \Theta $ with $\bm{\beta}$ fixed.
Meanwhile, the objective function $\mathcal{J}_{outer}$ optimizes $\mathcal{L}_{R}$ with respect to $ \bm{\beta} $ through $ \Theta^*(\bm{\beta}) $, considering $ \Theta^*(\bm{\beta})$ as a function of $\bm{\beta}$.
% Thus, we can have an alternating optimization to learn $ \Theta $ and $ \bm{\beta} $:
% \begin{itemize}
%     \item \textit{$ \Theta $ update phase} (Inner Optimization): Fix $ \bm{\beta} $ and optimize $ \Theta $.
%     \item \textit{$ \bm{\beta} $ update phase} (Outer Optimization): Fix $ \Theta $ and optimize $ \bm{\beta} $.
% \end{itemize}

\begin{algorithm}[hbt]
\SetAlgoLined
Initialize optimizers $\operatorname{OPT}_{\Theta}$ and $\operatorname{OPT}_{\bm{\beta}}$ \;
\While{not converged}{
$\Theta$ Update~(fix $\bm{\beta}^t$):\\ 
\Indp $\Theta^{t+1} \longleftarrow \operatorname{OPT}_{\Theta}\left(\Theta^t, \nabla_{\Theta^t} \mathcal{J}_{inner}(\Theta^t, \bm{\beta}^t)\right)$ \;
\Indm Proxy:\\
\Indp $\tilde\Theta^{t+1}(\bm{\beta}^t) := \Theta^t - \alpha \nabla_{\Theta^t} \mathcal{J}_{inner}(\Theta^t, \bm{\beta}^t)$ \;
\Indm$\bm{\beta}$ Update~(fix $\Theta^t$):\\
\Indp $\bm{\beta}^{t+1} \longleftarrow \operatorname{OPT}_{\bm{\beta}} \Big( \bm{\beta}^t, \nabla_{\bm{\beta}^t}\mathcal{J}_{outer}\big(\tilde\Theta^{t+1}(\bm{\beta}^t)\big)\Big)$ \;  
}
\caption{Iterative Training Procedure}
\label{alg:opt}
\end{algorithm}

As most existing models use gradient-based methods for optimization, a simple approximation strategy with less computation is introduced as follows:
\begin{equation}
    \begin{aligned}
    \nabla_{\bm{\beta}} \mathcal{J}_{outer}\left(\Theta^*(\bm{\beta})\right)\approx
    \nabla_{\bm{\beta}} \mathcal{J}_{outer} \left(\Theta - \alpha \nabla_{\Theta} \mathcal{J}_{inner}(\Theta, \bm{\beta}) \right)\,.
    \end{aligned}
\end{equation}
In this expression, $\alpha$ is the learning rate for one step of inner optimization. Related approximations have been validated in~\cite{DBLP:conf/wsdm/Rendle12,DBLP:conf/iclr/LiuSY19,DBLP:conf/kdd/MaMZTLC20}. Thus, we can define a proxy function to link $\bm{\beta}$ with the outer optimization:
\begin{equation}
    \tilde{\Theta}(\bm{\beta}) := \Theta - \alpha \nabla_{\Theta} \mathcal{J}_{inner}(\Theta, \bm{\beta})\, .
    \label{eq:gradient_approx}
\end{equation}
For simplicity, we use two optimizers $ \operatorname{OPT}_{\Theta} $ and $ \operatorname{OPT}_{\bm{\beta}} $ to update $ \Theta $ and $ \bm{\beta} $, respectively. The iterative procedure is shown in Alg.~\ref{alg:opt}.

\subsection{Training and Prediction}
After incorporating a parameter regularization term to avoid overfitting, the overall loss function is:
\begin{equation}
\begin{aligned}
    & \underset{\bm{\beta}}{\mathrm{min}} \: \mathcal{J}_{outer}\left(\Theta^*(\bm{\beta})\right) \\
    & \mathrm{s.t.} \: \Theta^{*}(\bm{\beta}) = \underset{\Theta}{\mathrm{argmin}}\: \mathcal{J}_{inner}(\Theta, \bm{\beta}) + \lambda ||\Theta||_\mathrm{F} \,,
\end{aligned}
\label{eq:final_loss}
\end{equation}
where $ \lambda $ is a hyper-parameter. When minimizing the objective function, the partial derivatives with respect to all the parameters can be computed by gradient descent with back-propagation. We apply the Adam~\cite{DBLP:journals/corr/KingmaB14} algorithm to automatically adapt the learning rate during the learning procedure.

\textbf{Recommendation Phase}. For user $ u $, we compute the distance $  d_{\mathbb{B}}(\mathbf{u}, \mathbf{v}) $ between the user $ u $ and each item $ v $ in the dataset. Then the items that are not in the training set and have the shortest distances are recommended to user $ u $.

% \textcolor{blue}{Discussion about the number of parameters}

\section{Evaluation}
In this section, we first describe the experimental set-up. We then report the results of the conducted experiments and demonstrate the effectiveness of the proposed modules.

\subsection{Datasets}
The proposed model is evaluated on three real-world datasets from various domains with different sparsities: \textit{Amazon-book}, \textit{Last-FM} and \textit{Yelp2018}, which are fully adopted from~\cite{DBLP:conf/kdd/Wang00LC19}. The \textit{Amazon-book} dataset is adopted from the Amazon review dataset~\cite{DBLP:conf/www/HeM16} with the \textit{book} category, which covers a large amount of user-item interaction data, e.g., user ratings and reviews. The \textit{Last-FM} dataset is collected from \textit{last.fm} music website, where the tracks are viewed as the items. A subset of data from Jan. 2015 to Jun. 2015 is selected. The \textit{Yelp2018} dataset is adopted from the 2018 edition of the \textit{Yelp} challenge, where local businesses like restaurants and bars are viewed as the items.

All the above datasets follow the 10-core setting to ensure that each user and item have at least ten interactions. For Amazon-book and Last-FM, items are mapped into Freebase entities via title matching if there is a mapping available. For Yelp2018, the item knowledge from the local business information network (e.g., category, location, and attribute) is extracted as KG data. The data statistics after preprocessing are shown in Table \ref{tab:data_statistics}. 

For fair comparison, these three datasets in our experiments are exactly the same as those used in~\cite{DBLP:conf/kdd/Wang00LC19}. For each dataset, 80\% of interaction data of each user is randomly selected to constitute the training set, and we treat the remaining 20\% as the test set. From the training set, 10\% of interactions are randomly selected as validation set to tune hyper-parameters. The experiments are executed five times and the average result is reported.

\begin{table}[ht]
\centering
\caption{\label{tab:data_statistics}The statistics of the datasets.}
\begin{tabular}{l|r|r|r}
\hline 
 & Amazon-book & Last-FM & Yelp2018 \\
\hline
\#Users & 70,679 & 23,566 & 45,919 \\
\#Items & 24,915 & 48,123 & 45,538 \\
\#Interactions & 847,733 & 3,034,796 & 1,185,068 \\
\hline 
\#Entities & 88,572 & 58,266 & 90,961 \\
\#Relations & 39 & 9 & 42 \\
\#Triplets & 2,557,746 & 464,567 & 1,853,704 \\
\hline
\end{tabular}
% \vspace{-0.3cm}
\end{table}

% \begin{table*}[ht]
% \centering
% \caption{\label{tab:performance_comparison}The performance comparison of all methods in terms of \textit{Recall@10}, \textit{Recall@20}, \textit{NDCG@10}, and \textit{NDCG@20}. The best performing method is boldfaced. The underlined number is the second best performing method. * indicates the statistical significance for $ p <= 0.01 $ compared to the best baseline method based on the paired t-test.}
% \begin{tabular}{|l|cccc|cccc|cc|}
% \hline
% \multirow{2}{*}{Methods} & \multicolumn{4}{c|}{Amazon-book} & \multicolumn{4}{c|}{Last.fm} & \multicolumn{2}{c|}{yelp2018} \\ \cline{2-11}
% & R@10 & R@20 & N@10 & N@20 & R@10 & R@20 & N@10 & N@20 & R@10 & R@20 \\ \hline
% FM &   & 0.1345 &  & 0.0886 &  & 0.0778 &  & 0.1181  &   &   \\
% NFM & 0.0891 & 0.1366 & 0.0723 & 0.0913 &  & 0.0829 &  & 0.1214 &  & \\ 
% \hline
% CKE & 0.0875 & 0.1343 & 0.0705 & 0.0885 &  & 0.0736 &  & 0.1184 &  &  \\
% CFKG & 0.0769 & 0.1142 & 0.0603 & 0.0770 &  & 0.0723 &  & 0.1143 &  &  \\
% \hline
% RippleNet & 0.0883 & 0.1336 & 0.0747 & 0.0910 &  & 0.0791 &  & 0.1238 &  &  \\
% GC-MC & & 0.1316 &  & 0.0874 &  & 0.0818 &  & 0.1253 &  &  \\
% KGAT & 0.1017 & 0.1489 & 0.0814 & 0.1006 &  & 0.0870 &  & 0.1325 &  &  \\
% \hline
% Hyper-Know & \textbf{0.1062} & \textbf{0.1534} & \textbf{0.0855} & \textbf{0.1075} & \textbf{0.0677} & \textbf{0.0949} & \textbf{0.1218} & \textbf{0.1533} & \textbf{} & \textbf{}  \\
% Imp. & \% &  \% & \% & \% & \% & \% &  \% & \% & \% & \%  \\
% \hline 
% \end{tabular}
% % \vspace{-0.3cm}
% \end{table*}

\begin{table*}[ht]
\centering
\caption{\label{tab:performance_comparison}The performance comparison of all methods in terms of \textit{Recall@20} and \textit{NDCG@20}. The best performing method is boldfaced. The underlined number is the second best performing method. * indicates the statistical significance for $ p <= 0.01 $ compared to the best baseline method based on the paired t-test.}
\begin{tabular}{|c|c c| c c| c c c| c| c|}
\hline
& \textbf{FM} & \textbf{NFM} & \textbf{CKE} & \textbf{CFKG} & \textbf{RippleNet} & \textbf{GC-MC} & \textbf{KGAT} & \textbf{Hyper-Know} & \multicolumn{1}{l|}{\textbf{Improv.}} \\\hline
\multicolumn{10}{|c|}{Recall@20} \\
\hline
\textit{Amazon-book} & 0.1345 & 0.1366 & 0.1343 & 0.1142 & 0.1336 & 0.1316 & \underline{0.1489} & \textbf{0.1534}* & 3.23\% \\
% \hline
\textit{Last-FM} & 0.0778 & 0.0829 & 0.0736 & 0.0723 & 0.0791 & 0.0818 & \underline{0.0870} & \textbf{0.0949}* & 9.08\% \\
% \hline
\textit{Yelp2018} & 0.0627 & 0.0660 & 0.0657 & 0.0522 & 0.0664 & 0.0659 & \textbf{0.0712} & \underline{0.0683} & N/A\\
\hline
\multicolumn{10}{|c|}{NDCG@20} \\ 
\hline
\textit{Amazon-book} & 0.0886 & 0.0913 & 0.0885 & 0.0770 & 0.0910 & 0.0874 & \underline{0.1006} & \textbf{0.1075}* & 6.86\% \\
% \hline
\textit{Last-FM} & 0.1181 & 0.1214 & 0.1184 & 0.1143 & 0.1238 & 0.1253 & \underline{0.1325} & \textbf{0.1533}* & 16.70\% \\ 
% \hline
\textit{Yelp2018} & 0.0768 & 0.0810 & 0.0805 & 0.0644 & 0.0822 & 0.0790 & \underline{0.0867} & \textbf{0.0897}* & 3.46\% \\
\hline
\end{tabular}
% \vspace{-0.3cm}
\end{table*}

\subsection{Evaluation Metrics} \label{sec:metrics}
We evaluate all the methods in terms of \textit{Recall@K} and \textit{NDCG@K}. For each user, Recall@K (R@K) indicates what percentage of her rated items emerge in the top $ K $ recommended items. NDCG@K (N@K) is the normalized discounted cumulative gain at $ K $, which takes the position of correctly recommended items into account. % $ K $ is set to $ 10 $.

\subsection{Methods Studied}
To demonstrate the effectiveness of our model, we compare to the following recommendation methods:
\begin{itemize}
\item \textbf{FM}~\cite{DBLP:conf/icdm/Rendle10}, a classical factorization model, which incorporates the second-order feature interactions between input features.
\item \textbf{NFM}~\cite{DBLP:conf/sigir/0001C17}, a state-of-the-art factorization model, which subsumes FM under a neural network.
\item \textbf{CKE}~\cite{DBLP:conf/kdd/ZhangYLXM16}, a representative regularization-based method, which exploits semantic embeddings derived from TransR~\cite{DBLP:conf/aaai/LinLSLZ15} to enhance the matrix factorization.
\item \textbf{CFKG}~\cite{DBLP:journals/algorithms/AiACZ18}, a model that applies TransE~\cite{DBLP:conf/nips/BordesUGWY13} on the unified graph including users, items, entities, and relations, casting the recommendation task as the prediction of (u, Interact, i) triplets.
% \item \textbf{MCRec}~\cite{DBLP:conf/kdd/HuSZY18}, a path-based model, which extracts qualified meta-paths as connectivity between a user and an item.
\item \textbf{RippleNet}~\cite{DBLP:conf/cikm/WangZWZLXG18}, a model that combines regularization- and path-based methods, which enrich user representations by adding those of items within paths rooted at each user.
\item \textbf{GC-MC}~\cite{DBLP:journals/corr/BergKW17}, a model designed to employ a graph convolutional network on graph-structured data. Here the model is applied on the user-item knowledge graph.
\item \textbf{KGAT}~\cite{DBLP:conf/kdd/Wang00LC19}, a state-of-the-art KG enhanced model, which employs a graph neural network and an attention mechanism to learn from high-order graph-structured data for recommendation.
\item \textbf{Hyper-Know}, the proposed model, which learns the knowledge-enhanced recommendation in the Poincaré ball and applies hyperbolic attention for distinguishing neighboring entities and bilevel optimization for adaptive regularization, respectively.
\end{itemize}

\subsection{Experiment Settings}
In the experiments, the latent dimension of all the models is set to 64. The parameters for all baseline methods are initialized as in the corresponding papers, and are then carefully tuned to achieve optimal performances. The learning rate is tuned amongst $ [0.0001, 0.0005, 0.001, 0.005, 0.01] $, and we search for the coefficient of L2 normalization over the range $ [0.0001 ,..., 0.1] $. To prevent overfitting, the dropout ratio is selected from the range $ [0.0, 0.1, ..., 0.9] $ for NFM, GM-MC, and KGAT. The dimension of attention network $ k $ is tested over the values $ [16, 32, 64] $. Regarding NFM, the number of MLP layers is set to $1$ with $64$ neurons according to the original paper.
For RippleNet, we set the number of hops and the memory size as $2$ and $8$, respectively. For KGAT, we set the depth as $3$ with hidden dimension $64$, $32$, and $16$, respectively. The network architectures of the above methods are configured to be the same as described in the original papers. For Hyper-Know, the curvature $ c $ is set to 1 and the batch size is set to $ 4096 $. The hyper-parameters are tuned on the validation set. Our experiments are conducted with PyTorch running on GPU machines (NVIDIA Tesla V100).

\subsection{Performance Comparison}
The performance comparison results are shown in Table \ref{tab:performance_comparison}. 

\textbf{Observations about our model}. 
First, the proposed model---Hyper-Know, achieves the best performance for most evaluation metrics on three datasets, which illustrates the superiority of our model. Second, Hyper-Know outperforms KGAT on the Amazon-book and Last-FM datasets. Although KGAT adopts the attention model to distinguish the entity importance in the knowledge graph, it may not effectively capture the hierarchical structure between entities, which can be well-modeled by learning the entity and relation embeddings in the hyperbolic space. One possible reason why Hyper-Know does not outperform KGAT for the Recall@20 metric on the Yelp2018 dataset is that most of the entities in this KG are linked according to whether they have the same attributes, such as \textit{HasTV}. Most of these attributes are very generic, which means that the KG provides information of limited value. As a result, much of the transfer that Hyper-Know performs from the KG to the recommendation part for the Yelp2018 dataset is likely to be noise. 
Third, Hyper-Know achieves better performance than GC-MC and RippleNet. Although GC-MC and RippleNet can model high-order connectivities, they fail to identify the important entities that would make a difference in recommendation. On the other hand, Hyper-Know employs an attention model in the hyperbolic space to learn the neighborhood representation of an item and transfers the knowledge from the KG to the item representation via regularization. 
Fourth, Hyper-Know obtains better results than CKE. One possible reason is that CKE adopts a fixed power of regularization during the whole training process. By contrast, Hyper-Know performs fine-grained regularization to regularize the item and its neighborhood. 
Fifth, Hyper-Know outperforms FM and NFM. One reason may be that using a distance as the scoring function can capture more fine-grained user preference.

\textbf{Other observations}. First, KGAT outperforms GC-MC and RippleNet. KGAT is capable of exploring the high-order connectivity in an explicit way and applies a graph attention model to aggregate the neighbors in the user-item knowledge graph in a weighted manner. Second, FM and NFM achieve better performance than CFKG and CKE in most cases. One major reason is that FM and NFM capture the second-order connectivity between users and entities, whereas CFKG and CKE model connectivity on the granularity of triples, leaving high-order connectivity untouched. Third, RippleNet achieves better performance than FM. This may verify that incorporating two-hop neighboring items is of importance to enrich user representations. Fourth, NFM performs better than FM. One major reason is that NFM has stronger expressiveness, since the hidden layer allows NFM to capture the nonlinear and complex feature interactions between user, item, and entity embeddings.

\begin{table}[ht]
\centering
\caption{\label{tab:ablation_analysis}The ablation analysis. \textit{Att} denotes the attention model, \textit{Avg} denotes the embedding average operation, \textit{E} denotes the Euclidean space, and \textit{H} denotes the hyperbolic space.}
\begin{tabular}{ |l|c|c|c|c| }
\hline
\multirow{2}{*}{Architecture} & \multicolumn{2}{c|}{\textit{Amazon-book}} & \multicolumn{2}{c|}{\textit{Last-FM}} \bigstrut \\\cline{2-5} 
& R@20 & N@20 & R@20 & N@20 \bigstrut \\ 
\hline
(1) BPR+E & 0.1017 & 0.0729 & 0.0604 & 0.1112 \\
(2) BPR+H & 0.1167 & 0.0833 & 0.0656 & 0.1191 \\
(3) BPR+Att+E & 0.1121 & 0.0812 & 0.0746 & 0.1319 \\
(4) BPR+Att+H & 0.1447 & 0.1025 & 0.0885 & 0.1453 \\
(5) BPR+Avg+H & 0.1250 & 0.0897 & 0.0775 & 0.1358 \\
(6) Hyper-Know & \textbf{0.1534} & \textbf{0.1075} & \textbf{0.0949} & \textbf{0.1533} \\
\hline
\end{tabular}
% \vspace{-0.3cm}
\end{table}

\subsection{Ablation Analysis} \label{sec:ablation}
To verify the effectiveness of the proposed model in the Poincaré ball, the hyperbolic attention model, and the adaptive regularization mechanism, we conduct an ablation study in Table~\ref{tab:ablation_analysis}. This demonstrates the contribution of each module to the Hyper-Know model. In (1), we use the Euclidean distance to measure the user preference optimized by the BPR loss. In (2), we apply the distance in the Poincaré ball to measure users' preferences and optimize using Eq.~\ref{eq:bpr_dist}. In (3), we integrate the TransE-style attention on the top of (1) in the Euclidean space. In (4), we add hyperbolic attention to (2). In (5), we replace the attention model in (4) with an average operation in the hyperbolic space. In (6), we present the overall Hyper-Know model to show the effectiveness of the adaptive regularization mechanism.

From the results shown in Table~\ref{tab:ablation_analysis}, we make the following observations. First, comparing (1) and (2), we can observe that measuring the user preference by calculating distance in the hyperbolic space achieves better performance than calculating distance in the Euclidean space. This confirms the results reported in~\cite{DBLP:conf/wsdm/TranT0CL20}. Second, from (2) and (3), we observe that incorporating the hyperbolic attention model significantly improves the model performance. Third, in (3) and (4), we compare the performance of the attention model in both the Euclidean and hyperbolic space. From the results, we can observe that the attention model achieves better results in the hyperbolic space than in the Euclidean space. Fourth, from (1), (2), (3) and (4), we can observe that equipping the recommendation model with the KG either in the Euclidean space or hyperbolic space can improve the recommendation performance. Fifth, from (4) and (5), we observe that by distinguishing the importance of each neighbour of an item through attention, we achieve considerable improvement compared to a simple average. Comparing (4) and (6), we can observe that the adaptive regularization can provide the fine-grained regularization power.

\begin{table}[ht]
\caption{\label{tab:training_time}Training time comparison.}
\begin{tabular}{|c|c|c|c|c|}
\hline
 & CKE & CFKG & KGAT & Hyper-Know \\ \hline
Amazon-book & 55s & 22s & 457s & \textbf{15s} \\ \hline
Last-FM & 53s & 27s & 137s & \textbf{22s} \\ \hline
Yelp2018 & 63s & 37s & 352s & \textbf{20s} \\
\hline
\end{tabular}
% \vspace{-0.3cm}
\end{table}

\subsection{Training Efficiency}
In this section, we compare the training efficiency with other state-of-the-art KG-enhanced methods in terms of the training speed. We compare the time taken for one epoch of training. 
%{\color{red} [** THIS IS ONLY FAIR/MEANINGFUL IF THE NUMBER OF EPOCHS REQUIRED TO CONVERGE IS APPROXIMATELY THE SAME - CAN YOU CONFIRM THAT THIS IS THE CASE?]} \textcolor{blue}{From the Table 5 of this paper--Jointly Non-Sampling Learning for Knowledge Graph Enhanced Recommendation, they approximate take the same epoch to converge. We almost take the similar epochs to converge.} 
From the results reported in~\cite{DBLP:conf/sigir/ChenZMLM20}, these compared methods take a similar number of epochs to converge as well as our proposed method.
RippleNet is not computation-efficient and takes much longer to train, we omit the comparison with RippleNet. All the experiments are conducted on a single GPU of an NVIDIA Tesla V100. All the compared methods are executed for 20 epochs and we report the average computation time, which is shown in Table~\ref{tab:training_time}. The training time comparison shows that Hyper-Know is more computationally efficient than other state-of-the-art methods, and the reason follows.
%{\color{red}[** CAN YOU EXPLAIN WHY?]} \textcolor{blue}{Compare to CKE, we have fewer learnable parameters; As for CFKG and KGAT, they incorporate the users into the knowledge graph, which substantially increase the number of nodes in KG.}
Compared to CKE, Hyper-Know has a smaller number of learnable parameters (8.3 million v.s. 11.4 million on the Last-FM dataset). Compared to KGAT and CFKG, Hyper-Know does not incorporate the users into the KG, which makes the scale of the KG much smaller.

\begin{figure}[ht!]
    \centering
    \begin{subfigure}[t]{0.25\textwidth}
        \centering
        \includegraphics[width=\linewidth]{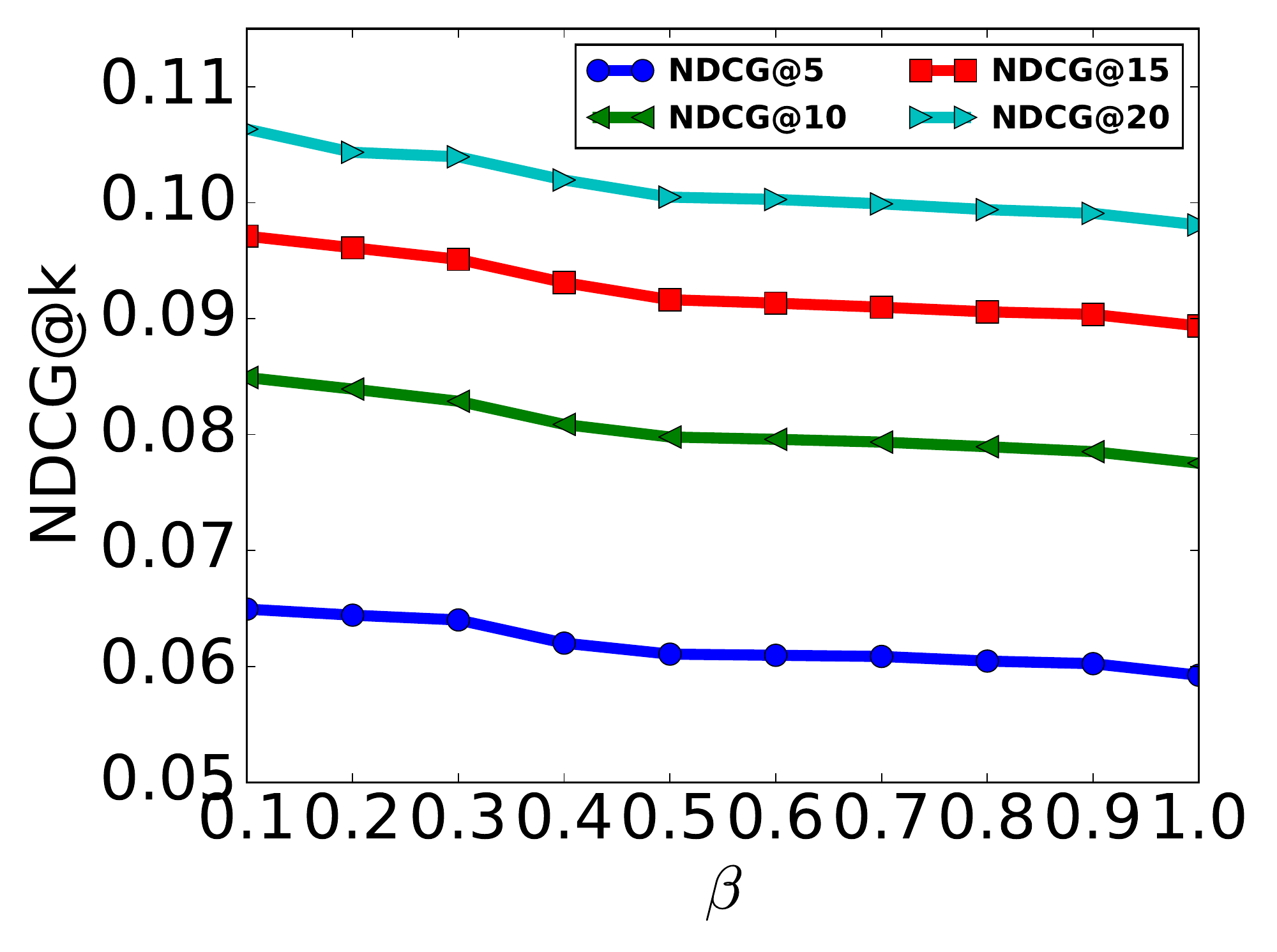}
        \caption{\label{fig:amazon_beta_var}$ \beta $ on Amazon-book}
    \end{subfigure}% 
    \begin{subfigure}[t]{0.25\textwidth}
        \centering
        \includegraphics[width=\linewidth]{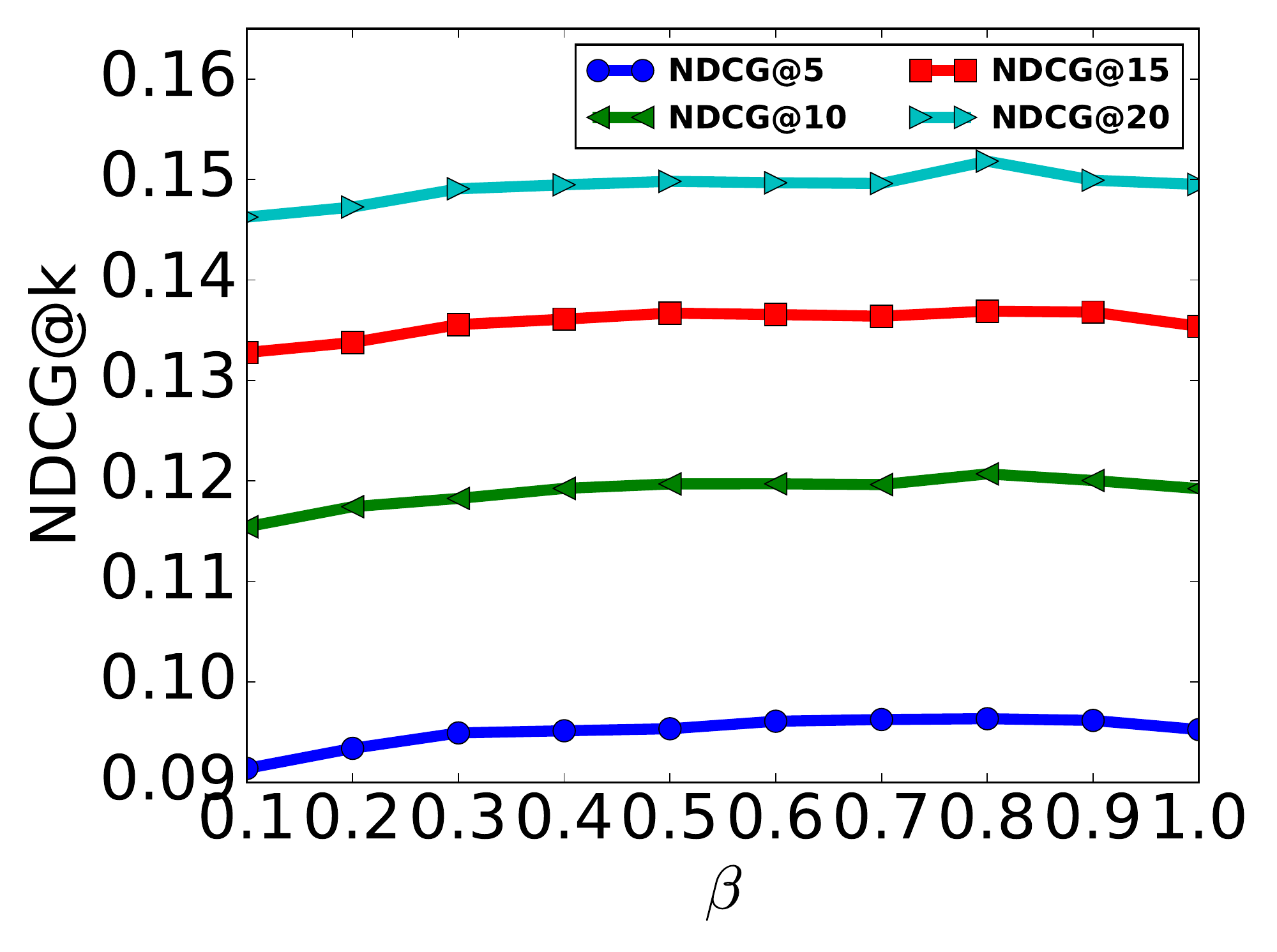}
        \caption{\label{fig:lastfm_beta_var}$ \beta $ on Last-FM}
    \end{subfigure}
    \caption{\label{fig:hyper_parameter}The variation of $ \beta $.}
% \vspace{-0.3cm}
\end{figure}

\subsection{Influence of Hyper-parameters}
The value of $ \beta $ that regularizing the item embedding with its neighborhood is an important hyper-parameter if not using the adaptive regularization mechanism. Its effect on the Amazon-book and Last-FM datasets is shown in Figure~\ref{fig:hyper_parameter}.

From the results in Figure~\ref{fig:hyper_parameter}, we observe that the value of $ \beta $ does affect the recommendation performance, with performance deteriorating by as much as 10 percent if a suboptimal value is chosen. Furthermore, there is no fixed value that achieves performance that is as good as the performance obtained by using the proposed adaptive mechanism. These results demonstrate that the fine-grained and adaptive regularization benefits the recommendation task, which confirms the results reported in~\cite{DBLP:conf/wsdm/Rendle12}.

% \begin{figure}[h]
%     \centering
%     \includegraphics[width=0.5\textwidth]{pic/heat_map.pdf}
%     \caption{The attention visualization of memory networks.}
%     \label{fig:heat_map}
% \vspace{-0.3cm}
% \end{figure}

\begin{figure}[ht!]
    \centering
    \begin{subfigure}[t]{0.2\textwidth}
        \centering
        \includegraphics[width=\linewidth]{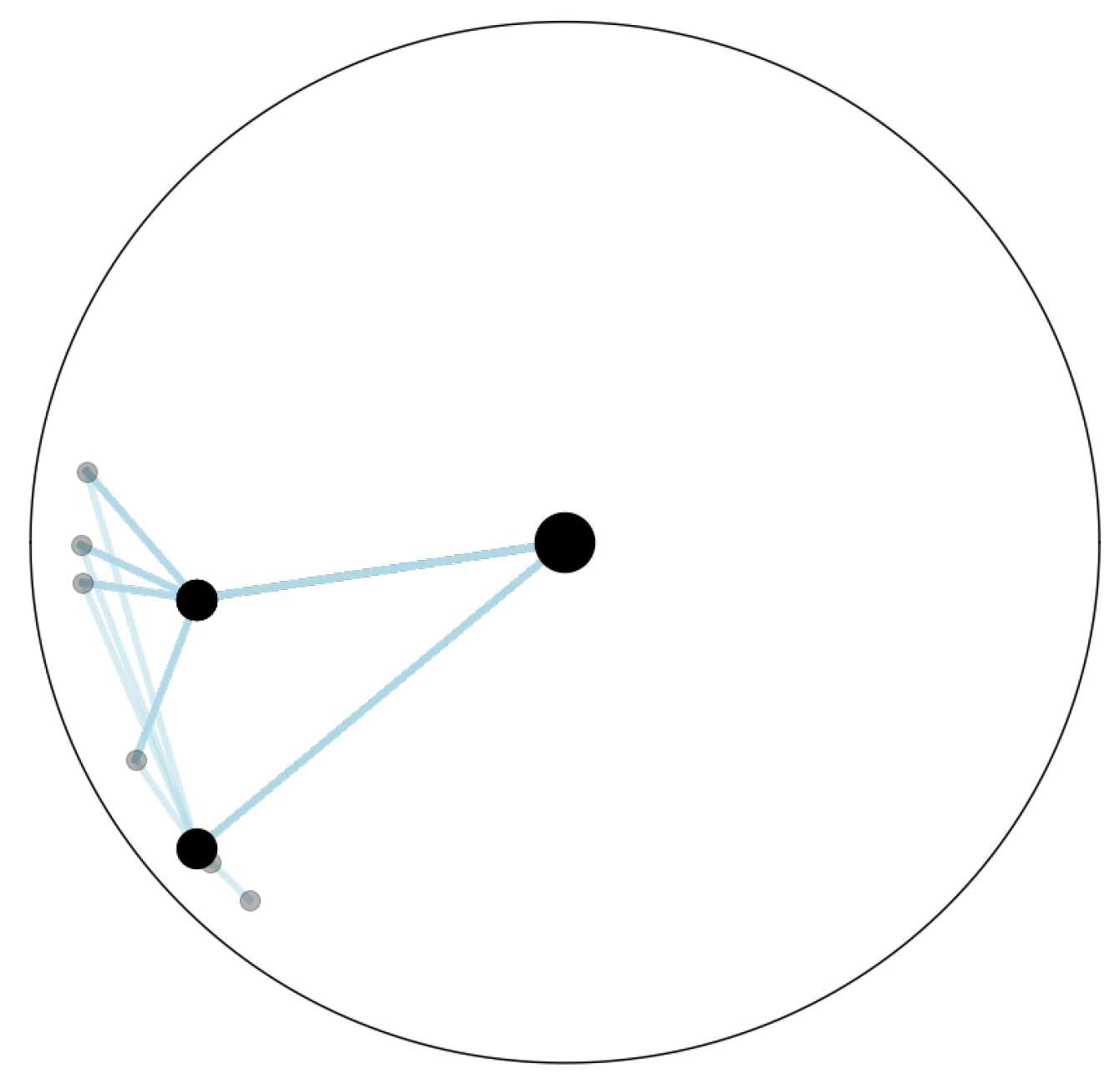}
        \caption{\label{fig:1}Entity-48130}
    \end{subfigure}% 
    \begin{subfigure}[t]{0.1\textwidth}
    \end{subfigure}
    \begin{subfigure}[t]{0.196\textwidth}
        \centering
        \includegraphics[width=\linewidth]{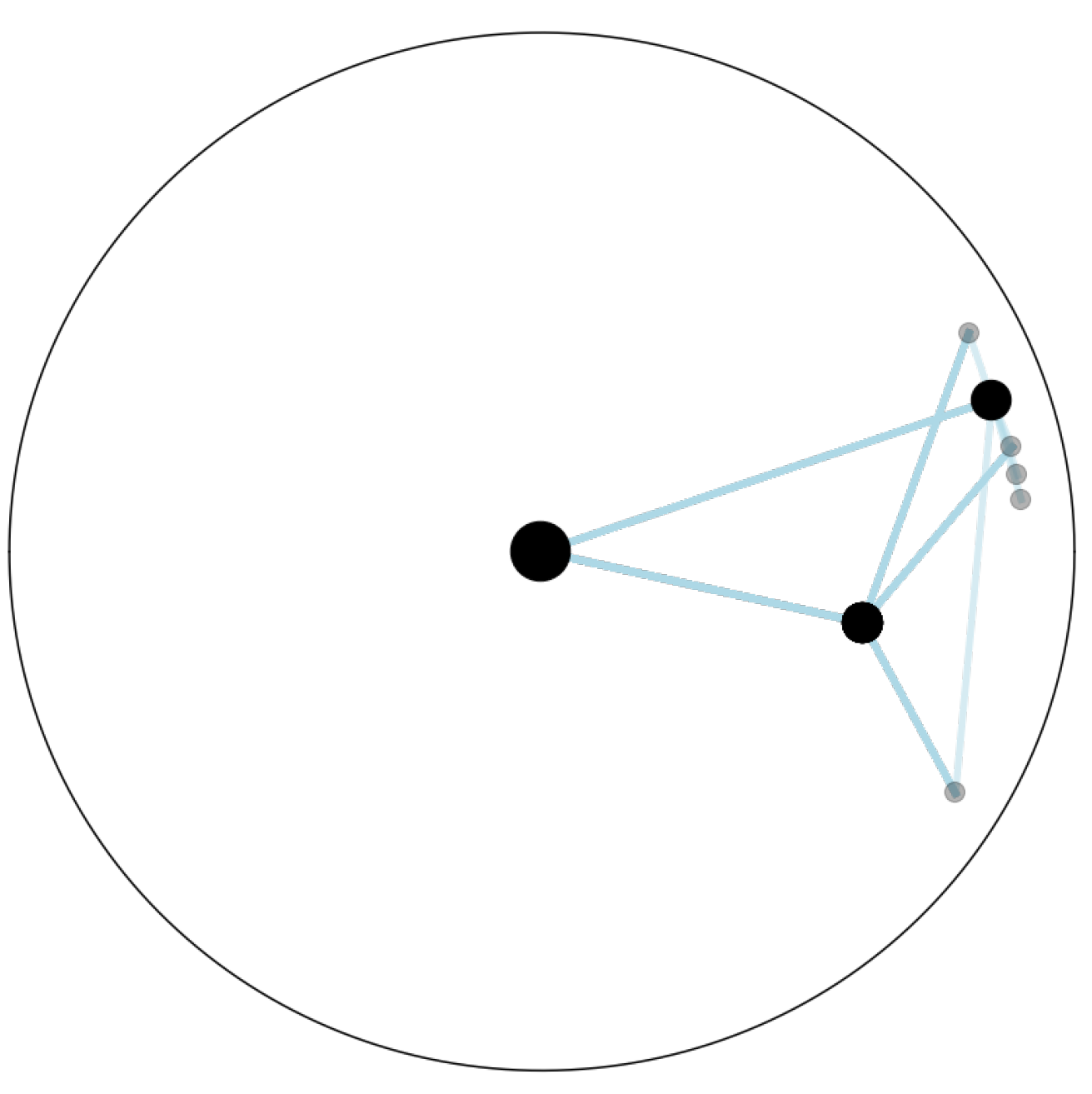}
        \caption{\label{fig:2}Entity-97468}
    \end{subfigure}
    \caption{\label{fig:case_study}The embedding visualization of selected entities.}
% \vspace{-0.3cm}
\end{figure}

\subsection{Embedding Visualization}
To verify whether the learned embedding in the Poincaré ball can capture the hierarchical structure in the knowledge graph, we train Hyper-Know in the 2D space on the Last-FM dataset and visualize the entities in the 2D hyperbolic space. We randomly select two nodes and their two-hop neighbors to visualize. 
The visualization is shown in Figure~\ref{fig:case_study}. The biggest dot denotes the selected entity, the less biggest dot denotes the first-hop neighbor of the selected entity, and the smallest dot denotes the second-hop neighbor.

From Figure~\ref{fig:case_study}, we can observe that these three kinds of nodes may form the hierarchical patterns in Poincaré ball and the learned embeddings in the hyperbolic space can represent the hierarchical relationships.

\section{Conclusion}
In this paper, we propose a knowledge-enhanced recommendation model in the hyperbolic space (Hyper-Know) for top-K recommendation. Hyper-Know learns the user and item embeddings as well as the knowledge graph representation in the Poincaré ball model to capture the hierarchical structure in the knowledge graph. In addition, we incorporate hyperbolic attention to select the most important neighboring entities of each item. To adaptively control the regularization effect, a bilevel optimization mechanism is proposed to generate a fine-grained regularization effect between recommendation and the knowledge graph. Experimental results on three real-world datasets clearly validate the performance advantages of our model over multiple state-of-the-art methods and demonstrate the effectiveness of each of the proposed constituent modules.

\bibliography{references.bib}

\end{document}